\documentclass[aps,prl,showpacs,twocolumn]{revtex4}
\usepackage{amsmath,bm}
\usepackage[dvips]{graphicx}

\begin{document}

\title{Probability of anomalously large Bit-Error-Rate in long haul optical transmission}

\author{V. Chernyak$^a$, M. Chertkov$^b$, I. Kolokolov$^{b,c,d}$, and V. Lebedev$^{b,c}$}

\affiliation{$^a$Corning Inc., SP-DV-02-8, Corning, NY 14831, USA; \\
$^b$Theoretical Division, LANL, Los Alamos, NM 87545, USA; \\
$^c$Landau Institute for Theoretical Physics, Moscow, Kosygina 2, 117334, Russia; \\
$^{d}$Budker Institute of Nuclear Physics, Novosibirsk 630090, Russia.}

\date{\today}

\begin{abstract}

We consider a linear model of optical pulse transmission through fiber with birefringent disorder in
the presence of amplifier noise. Both disorder and noise are assumed to be weak, i.e. the average
bit-error rate (BER) is small. The probability of rare violent events leading to the values of BER
much larger than its typical value is estimated. We show that the probability distribution has a long
algebraic tail.

\end{abstract}


\maketitle

Optical fibers are widely used for transmission of information. In an ideal case, information
carried by pulses would be transmitted non-damaged. In reality, however, various
impairments lead to the  information loss. Noise generated by optical amplifiers and fiber
birefringence are the
two major impairments in high-speed fiber communications. The amplifier noise is short-correlated
in time, while the birefringence varies significantly along the optical line and is
practically frozen in time, since the characteristic temporal scale of such variations is long
compared to the signal propagation time through the entire fiber line. Coexistence of two
different sources of randomness characterized by two well-separated time scales is pretty common in
statistical physics of disordered systems. A classical example would be the glassy behavior
driven by short-correlated thermal noise in a system with frozen structural disorder (see, e.g.,
\cite{GlassReview} for review). An extremely non-Gaussian statistics of the observables is an
important feature of the disordered/glassy system. In this letter we demonstrate that strong
deviation from Gaussianity is also typical for optical fiber telecommunication systems.

Birefringent disorder is caused by weak random ellipticity of the fiber cross section.
Birefringence splits the pulse into two polarization components and it also leads to pulse
broadening \cite{79US,81Kam,86PW}. This effect known as polarization mode dispersion (PMD) have
been extensively studied experimentally \cite{78RU,81MSK,87BPW,87ACMD,96GGWP,00NJKG} and
theoretically \cite{88Pol,88PBWS,91PWN}. PMD is usually characterized by the so-called PMD vector
that was found to obey Gaussian statistics  \cite{88Pol,88PBWS,91PWN}. It was also shown, e.g. in
\cite{94OYSE}, that first order PMD compensation corresponding to cancellation of the PMD vector
on the carrier frequency is experimentally implementable. Higher-order generalizations of the PMD
vector (introduced to resolve  complex frequency dependence of the PMD phenomenon with higher
accuracy) as well as a suggestion on how to compensate for PMD in the higher orders have been also
discussed \cite{98Bul} and implemented experimentally, e.g. in \cite{99MK}. This works were focused
on separating effects caused by PMD from other potential impairments. Common wisdom hiding behind
this strategy says that one should start with evaluating the two impairments separately and then
estimate the joint effect on the optical telecommunication system performance taking them on equal
footing. In this letter we challenge this equal-footing approach. We show that the effects of
temporal noise and structural disorder on the overall system performance may not be separated
since the bit-error-rate (BER) strongly depends on a realization of birefringent disorder. Thus,
the probability distribution function (PDF) of BER and especially its tail corresponding to large
values of BER are the objects of prime interest and practical importance for describing the
probability of the system outage.

Our letter is organized as follows. We start with discussing the dynamical equations for an optical pulse
evolution in a birefringent fiber also influenced by the amplifier noise. BER produced
by the amplifier noise for a given realization of the birefringent disorder is analyzed first.
Then we describe the PDF of BER, found by averaging over many realizations of disorder. Some general
remarks conclude the letter.

The envelope of the electromagnetic field propagating through optical fiber in the linear regime (i.e. at
relatively low pulse intensity), which is subject to PMD distortion and amplifier noise, satisfies
the following equation \cite{79US,81Kam,89Agr}
 \begin{eqnarray}
 \partial_z{\bm\varPsi}-i\hat{\varDelta}(z){\bm\varPsi}
 -\hat{m}(z)\partial_t{\bm\varPsi}
 -id(z)\partial_t^2{\bm\varPsi}={\bm\xi}(z,t),
 \label{PMD} \end{eqnarray}
where $z$ is the position along the fiber, $t$ is the retarded time (measured in the reference
frame moving with the optical signal), $\bm\xi$ is the amplifier noise, and $d$ is the chromatic
dispersion coefficient. The envelope ${\bm\varPsi}$ is a two-component complex field where the
components stand for two polarization states of the optical signal. Birefringent disorder is
characterized by two random Hermitian $2\times2$ traceless matrices $\hat{\Delta}$ and $\hat{m}$
measuring fiber birefringence in the first- and second-order and corresponding to the first two
terms of the expansion in $\omega-\omega_0$ with $\omega$  and $\omega_0$ being the signal and
carrier frequencies respectively. The disorder is frozen at least on all the propagation related
time scales, i.e. the two matrices can be considered to be $t$-independent. The random matrix,
$\hat{\varDelta}(z)$, can be excluded from the consideration by means of the transformation to the
coordinate system rotating together with polarization state of the signal at the carrier
frequency: $\bm{\varPsi}\to\hat{V}\bm{\varPsi}$, $\bm{\xi}\to\hat{V}\bm{\xi}$ and $\hat{m}\to
\hat{V}\hat{m} \hat{V}^{-1}$. Here, the unitary matrix $\hat{V}(z)$ is the ordered exponential,
$T\exp[i\int_0^z\mathrm d z'\hat{\varDelta}(z')]$  defined as the solution of the equation,
$\partial_z\hat{V} =i\hat{\varDelta}\hat{V}$, with $\hat{V}(0)=\hat{1}$. Hereafter we use the
notations $\hat m$, $\bm\xi$ and $\bm\varPsi$ for the renormalized objects whereas the original
objects will be no more referred to. With this notation the equation for $\bm\varPsi$ has the same
form as Eq. (\ref{PMD}) if we set  $\hat\varDelta=0$. The solution of the renormalized equation
can be partitioned into a sum of a homogeneous contribution, insensitive to the additive noise and
the inhomogeneous one, $\bm\phi$ and $\bm\varphi$, respectively:
 \begin{eqnarray}
 \bm\varphi=\hat W(z)\bm\varPsi_0(t), \quad
 \bm\phi=\int_0^z\!\!\!\mathrm d z'\,
 \hat W(z) \hat W^{-1}(z') \bm\xi(z',t),
 \label{PMD4} \\
 \hat W(z)=\exp\left[i\!
 \int_0^z\!\!\!\mathrm dz'
 d(z')\partial_t^2\right]\!
 T\!\exp\left[\int_0^z\!\!\!\mathrm dz'
 \hat m(z')\partial_t\right]\!,
 \label{PDM3} \end{eqnarray}
where $\bm\varPsi_0$ describes the input pulse shape (at $z=0$).

We assume the optical system length $Z$ is much larger than the distance between the next-nearest
amplifier stations. (The stations are set to compensate for losses in the pulse intensity.)
Coarse-graining on the inter-amplifier scale allows treating amplification in the continuous
limit. Zero in average additive noise, $\bm\xi$, is the amplification leftover. The amplifier
noise has Gaussian statistics \cite{94Des} and its correlation time (set by quantum excitation
processes in amplifiers) is much shorter then the pulse width. Therefore, $\xi$-statistics is
fully determined by its pair correlation function
 \begin{equation}
 \langle \xi_\alpha(z_1,t_1)
 \xi^\ast_\beta(z_2,t_2)\rangle
 =D_\xi\delta_{\alpha\beta}
 \delta\left(z_1-z_2\right)\delta(t_1-t_2),
 \label{xixi} \end{equation}
where the coefficient $D_\xi$ characterizes the noise strength.

Averaging over birefringent disorder is of different nature: statistics here is collected over
different fibers or, alternatively, over different states of the same fiber collected over time.
(Birefringence is known to vary on a time scale essentially exceeding the pulse propagation time.)
The matrix $\hat{m}$ can be expanded in the Pauli matrices, $\hat{m}(z)=h_j(z) \hat{\sigma}_j$,
where $h_j$ is a real three-component field. The field is zero in average and short-correlated in
$z$. It enters the observables described by Eqs. (\ref{PMD4},\ref{PDM3}) in an integral form and,
according to the central limit theorem (see, e.g., \cite{Feller}),  can be treated as a Gaussian
random field described by the following pair correlation function
 \begin{equation}
 \langle h_i(z_1)h_j(z_2)\rangle
 =D_m\delta_{ij}\delta(z_1-z_2),
 \label{hh} \end{equation}
where $D_m$ characterizes the disorder strength. Note that even though the original $\hat{m}$
entering Eq. (\ref{PMD}) is anisotropic the isotropy of $h_j$ [implied in Eq. (\ref{hh})] is
restored as a result of  the $\hat{m}\to\hat{V}\hat{m}\hat{V}^{-1}$ transformation.

We discuss here the case of the so-called return-to-zero (RZ) modulation format when pulses
in a given frequency channel are well separated in $t$. Detection of a pulse at the fiber
output corresponding to $z=Z$ requires a measurement of the pulse intensity, $I$,
 \begin{equation}
 I=\int \mathrm dt\,G(t)
 \left|{\cal K}\,\bm\varPsi(Z,t)\right|^2 \,,
 \label{nnn} \end{equation}
where the function $G(t)$ is a convolution of the electrical (current) filter function with the
sampling window function (limiting the information slot). The linear operator ${\cal K}$ in Eq.
(\ref{nnn}) stands for an optical filter and may also incorporate a variety of engineering
``tricks" applied to the output signal $\bm\varPsi(Z,t)$. Ideally, $I$ takes a distinct value if
the bit is ``1" and is negligible if the bit is ``0". Both the noise and the disorder enforce $I$
to deviate from its ideal value. One declares the output signal to code $0$ or $1$ if the value of
$I$ is less or larger than the decision threshold $I_0$. The information is lost if the output
value of the bit differs from the input one. The probability of such event should be small (this
is a mandatory condition for a successful fiber line performance) i.e. both impairments typically
cause only small distortion to a pulse. Formally, this means: $D_\xi Z\ll1$, $D_mZ\ll1$, where both
the initial signal width and its amplitude are rescaled to unity.

Below we focus our analysis on the initially ``1'' (unitary) bit. We do not consider the evolution
of a zero bit since it does not contribute to the anomalously large values of BER, which we are
mainly interested to describe. The probability to loose the unitary bit is $B=\int_0^{I_0}\mathrm d
I\,P(I)$. Here $P(I)$ is the PDF of the signal intensity (\ref{nnn}) (that fluctuates due to the
noise $\bm\xi$) for the initial signal $\bm\varPsi_0$ corresponding to the bit ``1''. In
engineering practice BER is measured collecting the statistics over many initially identical
pulses. (As different pulses sense different realizations of the noise, this averaging over
different pulses is actually equivalent to the noise averaging.) Repeating the measurement of $B$
many times (each separated from the previous one by a time interval larger than the characteristic
time of the disorder variations or, alternatively, making measurements on different fibers) one
constructs the PDF of $B$, ${\cal S}(B)$. The PDF achieves its maximum at $B_0$, a typical value
of $B$. Even though the average distortion of a pulse caused by the noise and disorder is weak,
rare but violent events may substantially affect the optical system performance. The probability
of such rare events is determined by ${\cal S}(B)$ taken at $B\gg B_0$. Our further analysis is
focused on this PDF tail.

Mentioning the importance of accounting for an appropriate form of the linear operator ${\cal K}$, we
restrict our discussion to two possibilities, when the optical filtering is accompanied by an overall
time shift (this operation is usually called ``setting the clock") or by a compensation achieved
by insertion of an additional piece of fiber with adjustable birefringence. Optical filter is
required to separate different frequency channels. In addition the filter smoothes out an
otherwise strong impact on the pulse caused by amplifier noise temporal ultra-locality. ``Setting
the clock" procedure is formalized as, ${\cal K}_{cl}{\bm\varPsi}={\bm\varPsi} (t-t_{cl})$, where
$t_{cl}$ is an optimal time delay. The so-called first-order compensation means
 \begin{equation}
 {\cal K}_1{\bm \varPsi}=
 \exp\left[-\int_0^Z\mathrm dz\,\hat{m}(z)
 \partial_t\right]{\bm\varPsi},
 \label{first} \end{equation}
where the usual exponent (instead of the ordered one) enters Eq. (\ref{PDM3}). In the engineering
practice or laboratory experiments all the three strategies (along with some additional others) can
be applied simultaneously. Below we imply that the optical filter is always inserted. As far as
the ``setting the clock" and the first-order compensation procedures are concerned we intend to
compare those effects with the pure (no compensation) case.

The first step in our calculations is to find the value of $B$ for a given realization of
disorder. In this case the $\bm\varphi$-part of the envelope $\bm\varPsi$ is a constant (dependent
on the disorder) while the $\bm\phi$-contribution fluctuates. One obtains from Eqs.
(\ref{PMD4},\ref{xixi}) that $\bm\phi$ is a zero mean Gaussian variable, with the pair correlation
function $\langle\phi_\alpha(z,t_1)\phi^\ast_\beta(z,t_2)\rangle=D_\xi
z\delta_{\alpha\beta}\delta(t_1-t_2)$. Note that statistics of $\bm\phi$ is sensitive  neither to
the chromatic dispersion coefficient $d$ no to the birefringence matrix $\hat{m}$. Thus, averaging
over the noise statistics is reduced to a Gaussian path integral over $\bm\phi$. The inequality
$D_\xi Z\ll 1$ justifies the saddle-point evaluation of the integral and also allows to estimate
$\ln B$ as $\ln P(I_0)$. Thus, one finds that the product $D_\xi Z \ln B$ is a negative quantity
of order one, insensitive to the noise characteristics. This quantity depends on the initial signal
profile $\varPsi_0$, disorder $h_j$, integral chromatic dispersion $\int_0^Z d(z)dz$, and also on
the details of the measurement and compensation procedures.

We next analyze the dependence of $B$ on the birefringent disorder. The smallness of
$D_\xi Z$ means that even weak disorder can produce large deviations in $B$ from its typical
value $B_0$, $\ln B_0\sim-(D_\xi Z)^{-1}$, corresponding to $h_j=0$. Therefore $\ln(B/B_0)$ can be
represented as a series in $h_j$. The leading contribution is found upon expanding the ordered
exponential (\ref{PDM3}) [entering $\bm\varphi$ in accordance with Eq. (\ref{PMD4})] in a series in
$\bm h$, followed by substituting the result for $\bm\varphi$ into Eq. (\ref{nnn}),
and performing the saddle-point
calculations, aiming to find $P(I_0)$. The smallness of $D_\xi Z$ enables us to substitute
$\ln B$ by $\ln P(I_0)$. Then, in the second-order in $h$, one obtains
 \begin{eqnarray} &&
 \ln(B/B_0)=\Gamma/(D_\xi Z), \qquad
 \Gamma=\mu_1 H_3 +\mu_2 H_j^2
 \label{phihh} \\ &&
 +\mu'_2\int_0^Z\!\!\mathrm d z_1 \,\int_0^{z_1}\!\!\mathrm d z_2 \,
 \left[h_1(z_1)h_2(z_2)\!-\! h_2(z_1)h_1(z_2)\right],
 \nonumber \end{eqnarray}
where $H_j=\int_0^Z\mathrm d z\,h_j(z)$ and the initial pulse $\bm\varPsi_0$ is assumed to be
linearly polarized. The coefficients $\mu$ entering Eq. (\ref{phihh}) are sensitive to the initial
signal profile $\varPsi_0$, chromatic dispersion and to the measurement procedure. The
coefficients can be calculated only numerically, however, some conclusions based on the general
form of Eq. (\ref{phihh}) can be drown without this detailed calculations.

If no correction procedure is applied to the output signal then $\mu_2\sim1$, whereas $\mu_1$ and
$\mu'_2$ are corrections related to the temporal asymmetry of the pulse (produced by the optical
filter) and to the so-called the pulse chirp (generated by the dispersion term), respectively. If
no compensation is applied the leading role is played by the first term in Eq. (\ref{phihh}), and
one arrives at $\ln{\cal P}\approx -\Gamma^2/(2\mu_1^2 D_m Z)$. If the ``setting the clock"
procedure applies (with $t_{cl}$ dependent on $\bm h$: $t_{cl}= t_0-H_3$, with $t_0$ being the
optimal time shift without disorder), the first-order term in the right-hand side of Eq.
(\ref{phihh}) cancels out and the second-order term is dominated by $\mu_2(H_1^2+H_2^2)$.
According to Eqs. (\ref{PMD4},\ref{PDM3}) the statistics of $B$ depends on the integral dispersion
$\int_0^Z\mathrm dz\,d(z)$. In practice some special technical efforts (dispersion compensation)
are made to reduce it, so that usually this quantity is small. If the initial pulse is real (there
is no chirp) and if the integral dispersion is small, then $\mu'_2$ is negligible. In this case Eq.
(\ref{hh}) leads to the following PDF of $\Gamma$
 \begin{eqnarray}
 {\cal P}(\Gamma)
 =\frac{1}{2\mu_2 D_mZ}
 \exp\left(-\frac{\Gamma}{2\mu_2 D_m Z}\right) \,.
 \label{pdfmu2} \end{eqnarray}
If the first-order compensation scheme described by Eq. (\ref{first}) is used the first two terms
in the right-hand side of Eq. (\ref{phihh}) cancel out, $\mu_1=\mu_2=0$, and if in addition
$\mu'_2\neq0$ one arrives at
 \begin{eqnarray}
 {\cal P}(\Gamma)=(2D_mZ\mu'_2)^{-1}
 \cosh^{-1}\left(\frac{\pi\Gamma}{2\mu'_2 D_m Z}\right),
 \label{pdfmu3} \end{eqnarray}
replacing Eq. (\ref{pdfmu2}). Comparing Eqs. (\ref{pdfmu2},\ref{pdfmu3}) with the result for the
no-compensation case we conclude that the compensations lead to much narrower PDF. If $\mu'_2$ is
also $0$ (i.e. if the first order compensation is applied and the output signal is not chirped)
then higher-order terms in the expansion of $\Gamma$ in $h_j$ should be accounted for. This
case will be discussed elsewhere.

\begin{figure}
 \includegraphics[width=0.3\textwidth]{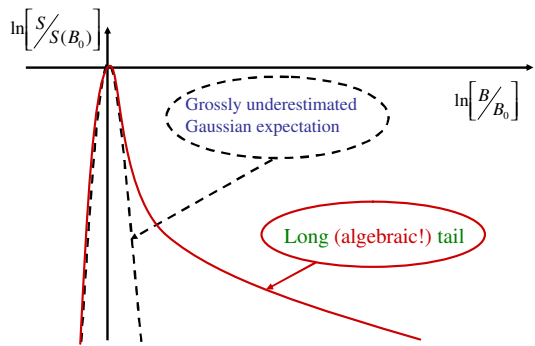}
 \caption{Schematic log-log plot of the PDF of Bit-Error-Rate.}
 \label{fig:scheme} \end{figure}

Finally, we are in a position to analyze the PDF of $B$, ${\cal S}(B)$. Expressing $\Gamma$ via $B$
according to (\ref{phihh}) followed by substituting the result into Eq. (\ref{pdfmu2}) for the PDF of
$\Gamma$ leads to the following expression for the remote (${\cal B}\gg {\cal B}_0$) tail of the
PDF of ${\cal B}$
 \begin{eqnarray} &&
 {\cal S}(B)\,\mathrm d B
 \sim\frac{B_0^\alpha\,\mathrm d B}
 {B^{1+\alpha}}, \qquad
 \alpha=\frac{D_\xi}{2\mu_2 D_m}.
 \label{power} \end{eqnarray}
The applicability range for Eq. (\ref{power}) is given by $1\gg\Gamma\gg D_mZ$ that transforms
into $1/(D_\xi Z)\gg\ln(B/B_0) \gg D_m/D_\xi$. If the first-order compensation is applied Eq.
(\ref{pdfmu2}) should be replaced by Eq. (\ref{pdfmu3}), thus leading to the same expression
(\ref{power}) for the tail with $\mu_2$ replaced by $\mu'_2/\pi$. The dependence of the PDF on BER
is illustrated in Fig. \ref{fig:scheme}. One also finds from Eq. (\ref{power}) that the so-called
outage probability, ${\cal O}\equiv\int^1_{B_\ast}\mathrm dB\,{\cal S}(B)$, where $B_\ast$ is some
fixed value taken to be much larger than $B_0$, is estimated as $\ln{\cal O}\sim(D_\xi/D_m)
\ln(B_0/B_\ast)$.

There is also a universal remote tail of ${\cal S}(B)$ corresponding to huge fluctuations of
the disorder when the signal is almost destroyed by the fluctuations. In this extreme case $I$ is
close to the threshold value $I_0$ already at $\bm\xi=0$. Then $B$ is nothing but a probability
of such an event (i.e. the noise configuration $\bm\xi$) that the inhomogeneous contribution
$\bm\phi$ would fill the remaining small gap between $I$ and $I_0$. Thus one gets $\ln B\sim
-\phi^2/(D_\xi Z)$. The value of $\bm\phi$ is proportional to the deviation $\delta h$ of the
$h$-disorder term from such a special configuration, $h_{sp}$, which gives $I=I_0$ at $\bm\xi=0$.
Therefore, one estimates, $\ln B\sim-(\delta h)^2 Z/D_\xi$. The logarithm of the probability for this
disorder configuration to occur is a sum of two contributions: $\sim - 1/(D_m Z)$, corresponding
to the special configuration, $h_{sp}$, and the other one, $\sim\delta h/D_m$, corresponding to
$\delta h$. One arrives at the following expression for the probability density
 \begin{eqnarray} &&
 \ln{\cal S}(B)\approx
 -\frac{C_1}{D_m Z}
 +C_2\sqrt{\frac{D_\xi}{D_m^2 Z}\ln\frac{1}{B}} \,,
 \label{remote} \end{eqnarray}
where $C_1$ and $C_2$ are constants of order one.
Eq. (\ref{remote}) holds when $(D_m Z)^2\ll D_\xi Z|\ln B|\ll 1$.
Note, that the remote tail (\ref{remote}) decays with $B$ faster
than the algebraic one (\ref{power}).

Eq. (\ref{power}) describes our major result: The PDF of BER has a long algebraic tail. The
exponent $\alpha$ of the algebraic decay is proportional to the ratio of the amplifier noise
variance, $D_\xi$, to the birefringent disorder variance, $D_m$. This statement clearly shows that
effects of noise and disorder are, actually, inseparable. Another interesting feature of Eq.
(\ref{power}) is that the exponent $\alpha$ is $Z$-independent. The only $Z$-dependent factor in
the final result (\ref{power}) is the overall normalization factor $B_0^\alpha$. Note also that
even though an extensive experimental (laboratory and/or field trial) confirmation of this
statement would be very welcome, some numerical results, consistent with Eq. (\ref{power}) are
already available. Thus, Fig. 2a of \cite{01XSKA} re-plotted in log-log variables shows a linear
relation between $\ln S$ and $\ln B$, i.e. just the algebraic decay predicted by Eq. (\ref{power}).

Summarizing, this letter is the first brief report on the new method/approach. We plan to discuss
many related but more specific issues, e.g. more complex compensation strategies (periodic and
higher-order compensation) in subsequent publications in specialized optical journals. However, it
is important to stress that the major result of the paper that the PDF of BER is characterized by a
long tail (algebraic or faster, but slower than exponential) is universal.

We are thankful to I. Gabitov for numerous extremely valuable discussions and G. Falkovich for
useful comments. We also wish to acknowledge the support of LDRD ER on ``Statistical Physics of
Fiber Optics Communications" at LANL.

\end{document}